# Scalable GWR: A linear-time algorithm for large-scale geographically weighted regression with polynomial kernels

Daisuke Murakami
Department of Statistical Modeling, The Institute of Statistical Mathematics,
10-3 Midori-cho, Tachikawa, Tokyo 190-8562, Japan
E-mail: dmuraka@ism.ac.jp

Narumasa Tsutsumida
Graduate School of Global Environmental Studies, Kyoto University,
Sakyo, Kyoto 606-8501, Japan

Takahiro Yoshida
Center for Global Environmental Research, National Institute for Environmental Studies. Tsukuba, Ibaraki 305-8506. Japan

Tomoki Nakaya
Graduate School of Environmental Studies, Tohoku University,
Sendai, Miyagi 980-0845, Japan

Binbin Lu
School of Remote Sensing and Information Engineering, Wuhan University,
Wuhan, Hubei 430079, China

**Abstract:** Although a number of studies have developed fast geographically weighted regression (GWR) algorithms for large samples, none of them has achieved linear-time estimation, which is considered a requisite for big data analysis in machine learning, geostatistics, and related domains. Against this backdrop, this study proposes a scalable GWR (ScaGWR) for large datasets. The key improvement is the calibration of the model through a pre-compression of the matrices and vectors whose size depends on the sample size, prior to the leave-one-out cross-validation, which is the heaviest computational step in conventional GWR. This pre-compression allows us to run the proposed GWR extension so that its computation time increases linearly with the sample size. With this improvement, the ScaGWR can be calibrated with one million observations without parallelization. Moreover, the ScaGWR estimator can be regarded as an empirical Bayesian estimator that is more stable than the conventional GWR estimator. We compare the ScaGWR with the conventional

GWR in terms of estimation accuracy and computational efficiency using a Monte Carlo simulation. Then, we apply these methods to a US income analysis. The code for ScaGWR is available in the R package **scgwr**. The code is embedded into C++ code and implemented in another R package, **GWmodel**.

## Introduction

In the last decade, in addition to the development of remote sensing, human sensing, and other sensing technologies, there has been a rapid increase in the size of spatial data. For example, more than 1 petabyte of spatial images are published by NASA per year whereas 10 million geo-tagged tweets are posted every day as of 2014 (Eldawy and Mokbel, 2016). In the era of open data, every researcher and practitioner can easily access high-resolution/large sensor data, survey data (e.g., census data), and voluntarily collected data (see Flanagin and Metzger, 2008).

Under this background, statistical methods for big spatial data have attracted considerable attention in geostatistics (for a review, see Heaton et al. 2018) and machine learning (for a review, see Liu et al. 2018), and other fields. In particular, a wide variety of geostatistical methods for very large samples have been developed (e.g., Cressie and Johannesson 2008; Banerjee et al. 2008; Finley 2011; Datta et al. 2016) and applied to analyze the spatial processes underlying carbon concentration, biomass, temperature, and other types of observations. Most of these methods have been designed to achieve "linear-time" computation (CP) complexity $O(N)$ with respect to the sample size $N$, where $O(\cdot)$ indicates the order. In other words, the computational cost doubles when the sample size doubles whereas the classical geostatistical interpolation method (i.e., kriging; see Cressie 1993) requires eight times the computational cost (i.e., $O(N^3)$). Today, linear-time CP is regarded as a necessary condition for modeling big spatial data.

Another important aspect of spatial big data is heterogeneity across samples. In other words, larger samples, which imply sampling of denser or wider areas, allow for further exploration of the local heterogeneity. For instance, US tract-level medium incomes, a large sample we will analyze later, vary depending on convenience, living environment, and other local factors. In contrast, it is almost impossible to analyze such local factors using state-level income data. Thus, the assumption of spatial homogeneity/stationarity, which has been assumed in geostatistics, is not appropriate for large spatial data. Modeling spatial heterogeneity is the key to analyzing big spatial data. Spatially varying coefficients (SVCs) modeling is a popular way to analyze spatial heterogeneity (e.g., Fotheringham, Brunsdon, and Charlton 2002). The SVCs model is useful for analyzing large heterogeneous samples. However, as we will review below, SVCs models are computationally demanding and unsuitable for large spatial samples.

Given this background, we develop a linear-time algorithm for geographically weighted regression (GWR; Brunsdon, Fotheringham, and Charlton 1998; Fotheringham Brunsdon, and Charlton 2002). GWR is a well-known approach for modeling spatial heterogeneity through local modeling and estimating SVCs. Due to its flexibility and simplicity, GWR has been widely accepted in regional science (e.g., Cahill and Mulligan, 2007, Páez, Long, and Farber 2008), epidemiology (e.g., Nakaya et al. 2005), and environmental science (e.g., Dong, Nakaya, and Brunsdon 2018) to investigate spatial heterogeneity in regression coefficients. Despite its popularity, GWR's application to big spatial data is limited because of its computational complexity. Efficient GWR algorithms have been developed by Harris et al. (2010), Tran, Nguyen, and Tran (2016), and Li et al. (2019) through parallelization. Li et al. (2019) optimized the linear algebra in the GWR algorithm to reduce the computational complexity to $O(N^2 \log N)$, which implies that the computational burden grows in a quasi-quadratic manner with $N$. Nevertheless, to the best of the author's knowledge, no existing algorithm achieves linear-time GWR modeling.

Other computationally efficient SVCs modeling approaches include (i) the spatial expansion method (Casetti 1972), (ii) the bivariate spline-based approach (Mu, Wang, and Wang (2018), (iii) the Moran eigenvector approach (Murakami et al. 2017, Murakami and Griffith 2019a, b), and (iv) geostatistical approaches (e.g., Finley 2011). They estimate SVCs by fitting spatially smooth functions, defined using $L(\ll N)$ spatial basis functions describing principal map patterns. However, smooth function approaches are known to suffer from the degeneracy problem (Stein 2014). In other words, the existing approaches cannot capture local-scale spatial variations, and their SVCs estimates tend to have overly smooth map patterns. Unfortunately, this tendency becomes severe as $N$ increases (because of the fixed $L$).

In summary, an SVCs modeling approach that fulfills the following requirements is required:

(i) Linear-time estimation of SVCs
(ii) Avoidance of the degeneracy problem, that is, small-scale spatial variations should be effectively captured.

With its focus on (ii), GWR, a local approach that does not suffer from the degeneracy problem, is an appropriate approach. Thus, we extend GWR to scalable GWR (ScaGWR) to satisfy all the requirements above.

The remainder of this paper is organized as follows: The next section introduces the original GWR, and the subsequent section develops the ScaGWR. Then, we compare GWR and ScaGWR through Monte Carlo experiments. Subsequently, we apply ScaGWR to analyze US income data. We then conclude with a discussion.

## Geographically weighted regression (GWR)

The basic GWR model can be expressed as follows:

$$y_i = \beta_{i,0} + \sum_{k=1}^{K} x_{i,k} \beta_{i,k} + \varepsilon_i \quad \varepsilon_i \sim N(0, \sigma^2), \tag{1}$$

where $i \in \{1, \cdots, N\}$ is an index for the sample sites distributed across a geographical space, $y_i$ is the $i$-th explained variable, $x_{i,k}$ is the $k$-th covariate, $\beta_{i,0}$ is the intercept parameter, and $\beta_{i,k}$ is the local regression coefficient for the $k$-th covariate. The regression coefficients $\boldsymbol{\beta}_i = [\beta_{i,0}, \beta_{i,1} \cdots \beta_{i,K}]'$ are estimated by a weighted least squares method, where " ′ " denotes the matrix transpose. The estimator is given by:

$$\widehat{\boldsymbol{\beta}}_i = \left[\mathbf{X}' \mathbf{G}_i^{(0)}(h) \mathbf{X}\right]^{-1} \mathbf{X}' \mathbf{G}_i^{(0)}(h) \mathbf{y}, \tag{2}$$

where $\mathbf{y}$ is the vector of the dependent variables and $\mathbf{X}$ is the matrix of covariates that includes a column of 1s for the intercept. $\mathbf{G}_i^{(0)}(h)$ is a diagonal matrix whose $j$-th element is the geographical weight $g_{i,j}^{(0)}(h)$ for the $j$-th sample and can be calculated via a distance-decaying kernel function. For example, the Gaussian kernel in Eq. (3) is the usual choice:

$$g_{i,j}^{(0)}(h) = \exp\left[-\left(\frac{d_{i,j}}{h}\right)^2\right], \tag{3}$$

where $d_{i,j}$ is the Euclidean distance between sample sites $i$ and $j$. Instead of Euclidean distance, the Minkowski distance (Lu et al. 2016), network distance, travel time, or other non-Euclidean distances (Lu et al. 2014a) can be used. $h$ is a bandwidth parameter that takes a small (near-zero) value if the regression coefficients have small-scale map patterns and a large value when they have large-scale patterns. As $h$ grows, $\widehat{\boldsymbol{\beta}}_i$ asymptotically converges to the ordinary least squares (OLS) estimator $\widehat{\boldsymbol{\beta}}_{OLS} = (\mathbf{X}'\mathbf{X})^{-1}\mathbf{X}'\mathbf{y}$.

The bandwidth $h$ can be optimized by leave-one-out cross-validation (LOOCV), which minimizes the cross-validation (CV) score. This score is defined as follows:

$$\text{CV score} = \sum_{i=1}^{N} \hat{\varepsilon}_i^2, \tag{4}$$

where $\hat{\varepsilon}_i = y_i - \sum_{k=1}^{K} x_{i,k} \hat{\beta}_{-i,k}$. $\hat{\beta}_{-i,k}$ is estimated using $N-1$ observations remaining after the $i$-th observation has been excluded. Specifically, $\widehat{\boldsymbol{\beta}}_{-i} = [\beta_{-i,1} \cdots \beta_{-i,K}]'$ is estimated using Eq. (2), where $\mathbf{G}_i(h)$ is replaced with $\mathbf{G}_{-i}(h)$, which is a diagonal matrix whose elements are represented by the vector $\mathbf{g}_{-i}(h) = [g_{1,j}(h), \cdots g_{i-1,j}(h), 0, g_{i+1,j}(h), \cdots g_{N,j}(h)]'$.

Figure 1 summarizes the LOOCV procedure for the basic GWR. Matrices or vectors whose dimensions depend on the sample size are shown in grey. As illustrated by the figure, large matrices must be repeatedly processed to find the optimal $b$. This is the main reason GWR is slow for large $N$.

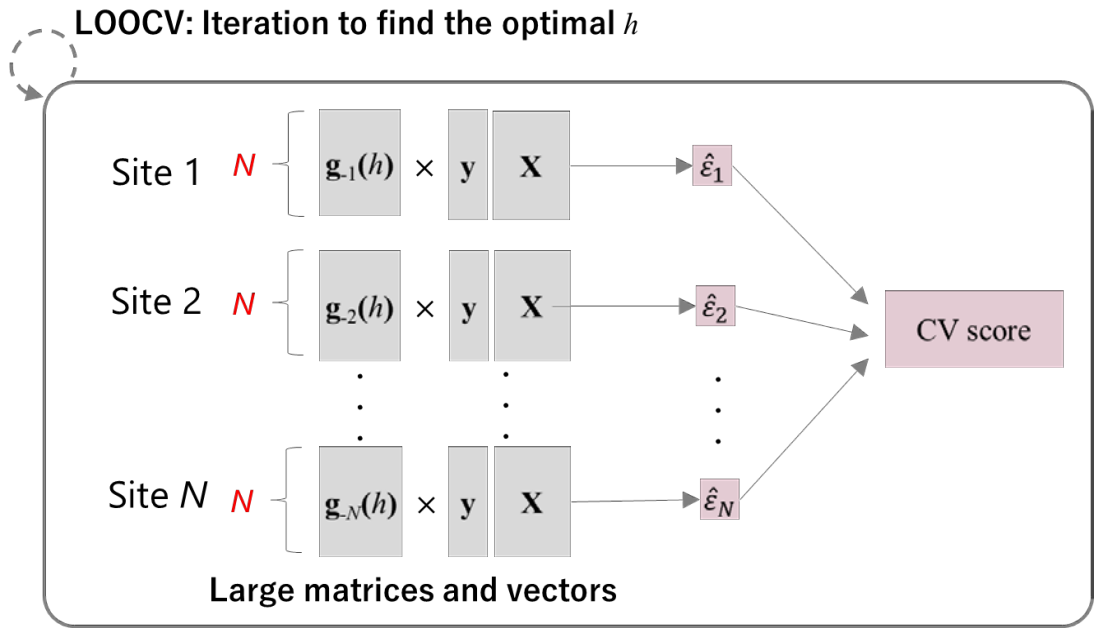

Figure 1. LOOCV routine in standard GWR. Grey squares represent matrices and vectors whose sizes depend on $N$, and red squares represent other small elements.

## Scalable geographically weighted regression (ScaGWR)

Figure 1 suggests that GWR can be accelerated if we let the large matrices/vectors out of the LOOCV iteration. However, this is not possible because the bandwidth parameter $h$ is not separable from $\mathbf{g}_{-i}(h)$. We overcome this bottleneck by introducing a linear multiscale kernel that allows us to linearly separate the bandwidth parameter from the kernel function and pre-process the large matrices and vectors before the LOOCV iterations. The resulting procedure estimates a regularized GWR model in quasi-linear time. We refer to this accelerated GWR as ScaGWR.

The subsections below explaining ScaGWR are organized as follows: The first subsection introduces a linear kernel that approximates standard kernels. The second subsection extends this linear kernel to a linear multiscale kernel for memory saving. The third subsection explains parameter estimation and the fourth describes the LOOCV, which is the critical step for fast CP.

### *Linear kernel*

Our kernel must be defined by a linear function to allow the large matrices to be excluded from the iterative part shown in Figure 1. Although a Taylor or Fourier transformation could be used to find a linear approximation of the kernels, our preliminary analysis suggests that their approximation error for the kernels rapidly increases with $d_{i,j}$. Instead, we propose the following linear kernel:

$$g_{i,j}^{(poly)}(b) = \sum_{p=1}^{P} b^p \, g_{i,j}(h_0)^{4/2^p}, \qquad (5)$$

where $g_{i,j}(h_0)$ is a base kernel with a known bandwidth $h_0$. We assume that $g_{i,j}(h_0)$ is a non-negative decreasing function with respect to $d_{i,j}$, and is equal to 1 if $d_{i,j} = 0$. $p \in \{1, \cdots, P\}$ represents the degree of the polynomials. This means that the Gaussian kernel (Eq. 3) and the exponential kernel can be used to define the $g_{i,j}(h_0)$ function whereas kernels with hard thresholds, such as the bi-square and tri-cube kernels, cannot. It is unclear how to determine $P$. Later, we perform a Monte Carlo simulation to clarify this.

Instead of the known $h_0$, we estimate $b$, which indicates the spatial scale of the SVCs. For instance, consider a case with three polynomials. In this case, Eq. (5) is the sum of the known base kernels $\{g_{i,j}(h_0)^2, g_{i,j}(h_0), g_{i,j}(h_0)^{1/2}\}$ weighted by the coefficients $\{b, b^2, b^3\}$. Among the base kernels, $g_{i,j}(h_0)^2$ decays the fastest whereas $g_{i,j}(h_0)^{1/2}$ decays the slowest with respect to distance. In other words, the three kernels describe the short-range, moderate-range, and long-range decay of a kernel function. The coefficients $\{b, b^2, b^3\}$ assign weights to these kernels. For example, if $b = 0.1$, the weights of the three kernels become $\{0.1, 0.01, 0.001\}$, with the largest weight assigned to the short-range kernel. The linear kernel resulting from Eq. (5) yields a short-range kernel. By contrast, if $b = 10$, the weights become $\{10, 100, 1000\}$, implying a long-range kernel. Thus, $b$ replaces the usual bandwidth parameter $h$. See Figure 2 for another example, with $P = 5$.

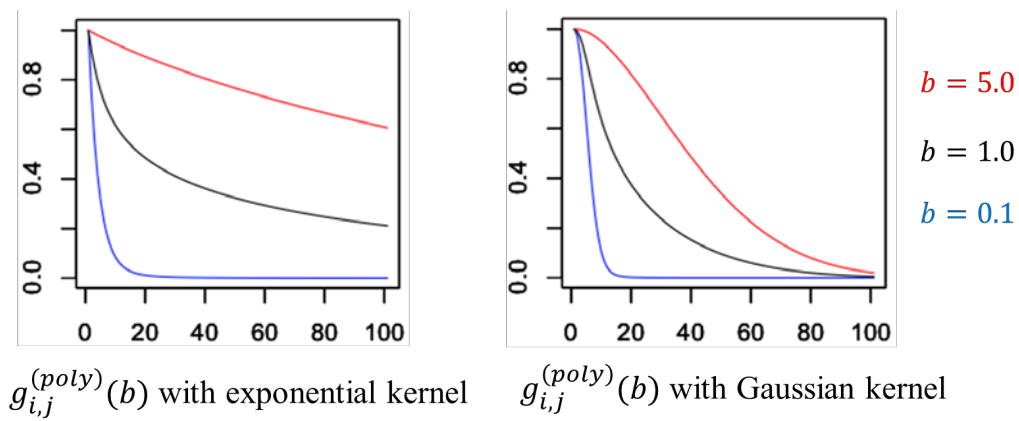

$g_{i,j}^{(poly)}(b)$ with exponential kernel $\qquad$ $g_{i,j}^{(poly)}(b)$ with Gaussian kernel

Figure 2. Example of a polynomial kernel with $P = 5$ and $b = 10$.

*Linear multiscale kernel*

The CP cost of calculating the linear kernel values for all the $N^2$ sample site pairs is $O(N^2)$. In other words, the use of the linear kernel implies a failure to achieve linear-time GWR estimation. To reduce the cost, we use a multiscale kernel, specified as

$$g_{i,j}(b, \alpha) = \alpha + \sum_{p=1}^{P} b^p\, g_{i,j}^{(Q)}(h_0)^{4/2^p}, \tag{6}$$

where

$$g_{i,j}^{(Q)}(h_0) = \begin{cases} g_{i,j}(h_0) & if\ d_{i,j} \le D_i^{(Q)} \\ 0 & otherwise \end{cases},$$

where $D_i^{(Q)}$ is the distance between the *i*-th site and the *Q*-th nearest neighbor. $\alpha$ and $b$ are parameters estimated by a LOOCV, which will be explained later, whereas $h_0$, $P$, and $Q$ are given a priori. The linear multiscale kernel is defined by the sum of a global weight ($\alpha$) and a local weight ($\sum_{p=1}^{P} b^p\, g_{i,j}(h_0)^{4/2^p}$) for the *Q*-nearest neighbors. The local weights are calculated only for the Q-nearest neighbors of each site. As a result, the complexity of calculating all $g_{i,j}(b, \alpha)$s is only $O(QN)$, which is trivial as long as $Q$ is small (e.g., 50). Furthermore, as we explain below, the resulting GWR estimator, which is known to suffer from multicollinearity (Wheeler and Tiefelsdorf 2005), is stabilized by using this kernel. For these reasons, we prefer the multiscale kernel. Note that a multiscale kernel is similar with an adaptive kernel (with hard thresholding) (see Fotheringham, Brunsdon, and Charlton 2002) in terms of assigning geographical weights only to the *Q*-nearest neighbors. However, unlike an adaptive kernel, a multiscale kernel assumes the same decay speed across the study area.

We need to determine the value of the fixed $h_0$ a priori. To capture local heterogeneity, the kernel must decay sufficiently within the *Q*-nearest neighbors. At the same time, to avoid a singular fit, the decay cannot be too fast. Given these points, the value of $b_0$ is defined so that the effective bandwidth (see Cressie 1993), which is the distance at which 95% of the kernel weight vanishes, equals the median of the $Q$-the nearest neighbor distances, $D_{med}^{(Q)}$. If $g_{i,j}(h_0)$ is the Gaussian kernel, the effective bandwidth equals $\sqrt{3}h_0$. Thus, we assume $h_0 = D_{med}^{(Q)}/\sqrt{3}$ (which is obtained by setting $\sqrt{3}h_0 = D_{med}^{(Q)}$). Similarly, we assume $D_{med}^{(Q)} = 3h_0$ for the exponential kernel. Because $b$ replaces $h_0$ as illustrated in Figure 2, the ScaGWR estimator is likely to be less sensitive to $h_0$ although a sensitivity analysis is an important future research topic.

*Coefficients estimator*

The ScaGWR model is identical to the standard GWR estimator, except for the kernel specification. Thus, the coefficients estimator becomes (see Eq. 2)

$$\widehat{\boldsymbol{\beta}}_i = [\mathbf{X}'\mathbf{G}_i(b,\alpha)\mathbf{X}]^{-1}\mathbf{X}'\mathbf{G}_i(b,\alpha)\mathbf{y}, \quad (7)$$

where $\mathbf{G}_i(b,\alpha)$ is a diagonal matrix whose $j$-th entry is $g_{i,j}(b,\alpha)$. By substituting Eq. (6) into Eq. (7), the estimator can be rewritten as

$$\widehat{\boldsymbol{\beta}}_i = [\alpha\mathbf{X}'\mathbf{X} + \mathbf{X}'\mathbf{G}_i^{(Q)}(b)\mathbf{X}]^{-1}[\alpha\mathbf{X}'\mathbf{y} + \mathbf{X}'\mathbf{G}_i^{(Q)}(b)\mathbf{y}], \quad (8)$$

where $\mathbf{G}_i^{(Q)}(b)$ is a diagonal matrix whose $j$-th entry is $\sum_{p=1}^{P} b^p g_{i,j}^{(Q)}(h_0)^{4/2^p}$. Based on Eq. (8), $\widehat{\boldsymbol{\beta}}_i$ can be viewed as an empirical Bayes estimator with prior distribution $\boldsymbol{\beta}^{prior} \sim N(\widehat{\boldsymbol{\beta}}_{OLS}, \alpha^{-1}(\mathbf{X}'\mathbf{X})^{-1})$. In other words, ScaGWR stabilizes the GWR estimator by shrinking it toward the OLS estimator, and $\alpha$ determines the degree of shrinkage. The OLS prior imposes a stronger regularization to the coefficients that take extreme values. Therefore, ScaGWR estimates are less likely to produce extreme or singular values whereas classical GWR often produces extreme coefficient estimates (e.g., Farber and Yeatest, 2006; Cho et al. 2009).

Once the parameters $\{b, \alpha\}$ are specified given, $\widehat{\boldsymbol{\beta}}_i$ is easily estimated. The next section describes our LOOCV procedure, which minimizes the CV score in Eq. (4) to optimize $\{b, \alpha\}$.

### *LOOCV for optimizing parameters $\{b, \alpha\}$*

To calculate the CV score for the LOOCV, we need to calculate $\widehat{\boldsymbol{\beta}}_{-i}$ for each *i* repeatedly:

$$\widehat{\boldsymbol{\beta}}_{-i} = [\mathbf{X}'\mathbf{G}_{-i}(b,\alpha)\mathbf{X}]^{-1}\mathbf{X}'\mathbf{G}_{-i}(b,\alpha)\mathbf{y}, \qquad (9)$$

where $\mathbf{G}_{-i}(b,\alpha)$ is equal to $\mathbf{G}_i(b,\alpha)$ with zero values in the *i*-th diagonal. We introduce matrix manipulation to make the calculation in Eq. (9) computationally efficient. The idea is to pre-process $\mathbf{X}'\mathbf{G}_{-i}(b,\alpha)\mathbf{X}$ and $\mathbf{X}'\mathbf{G}_{-i}(b,\alpha)\mathbf{y}$ before the LOOCV. As explained in Appendix 1, Eq. (9) is equivalent to the following matrix expression:

$$\widehat{\boldsymbol{\beta}}_{-i} = \left[\alpha\mathbf{M}_{-i}^{(0)} + \sum_{p=1}^{P} b^p \mathbf{M}_{-i}^{(p)}\right]^{-1} \left[\alpha\mathbf{m}_{-i}^{(0)} + \sum_{p=1}^{P} b^p \mathbf{m}_{-i}^{(p)}\right], \qquad (10)$$

where $\mathbf{M}_{-i}^{(0)}$ and $\mathbf{M}_{-i}^{(p)}$ are $K \times K$ matrices whose (*k*, *k'*)-th elements are $\sum_{j \neq i} x_{j,k} x_{j,k'}$ and $\sum_{j \neq i} g_{i,j}(h_0)^{4/2^p} x_{j,k} x_{j,k'}$, respectively, whereas $\mathbf{m}_{-i}^{(0)}$ and $\mathbf{m}_{-i}^{(p)}$ are $K \times 1$ vectors whose *k*-th elements are $\sum_{j \neq i} x_{j,k} y_j$ and $\sum_{j \neq i} g_{i,j}(h_0)^{4/2^p} x_{j,k} y_j$, respectively.

Importantly, because the $\mathbf{M}_{-i} \in \{\mathbf{M}_{-i}^{(0)}, \mathbf{M}_{-i}^{(p)}, \mathbf{m}_{-i}^{(0)}, \mathbf{m}_{-i}^{(p)}\}$ do not include any parameters, they can be calculated before the LOOCV. Given these elements, the computational complexity of calculating $\widehat{\boldsymbol{\beta}}_{-i}$ is only $O(K^3)$. Thus, the cost of estimating the coefficients for all the sample sites is $O(K^3N)$. If the golden section algorithm with an expected number of iterations of $O(\log(N))$ is used, and *K* is assumed as fixed, the complexity of the LOOCV is $O(N\log(N))$, which is a quasi-linear computational complexity with respect to $N$. This complexity is considerably smaller than that of fast GWR algorithms, whose complexity is $O(N^2\log(N))$ at best.

On the other hand, during the LOOCV, our algorithm needs to store $\mathbf{M}_{-i}$ for $i \in \{1, \cdots, N\}$, with a total number of $N(P + 1)(K^2 + K)$ elements. Thus, although the required memory for ScaGWR and fast GWR (Lu et al. 2019) are both linear with respect to $N$, ScaGWR consumes larger memory than fast GWR, whose memory consumption is of the order of $NK$. Reduction of memory usage is an important future research task.

The ScaGWR modeling procedure can be summarized as follows:

(I) Pre-processing: $\mathbf{M}_{-i}$ is calculated for each $i$.

(II) LOOCV: $\{b, \alpha\}$ are optimized by LOOCV, in which the CV score is calculated by substituting the inner products in $\mathbf{M}_{-i}$ into Eq. (10). As explained above, the complexity of this step is only $O(N\log(N))$ (see Figure 3).

(III) Estimation: $\widehat{\boldsymbol{\beta}}_i$ is estimated by substituting the estimated $\{b, \alpha\}$ into Eq. (11), which is equivalent to Eq. (8):

$$\widehat{\boldsymbol{\beta}}_i = \left[\alpha \mathbf{M}^{(0)} + \sum_{p=1}^{P} b^p \mathbf{M}_i^{(p)}\right]^{-1} \left[\alpha \mathbf{m}^{(0)} + \sum_{p=1}^{P} b^p \mathbf{m}_i^{(p)}\right], \quad (11)$$

where the $(k, k')$-th elements of the matrices $\mathbf{M}^{(0)}$ and $\mathbf{M}_i^{(p)}$ are given by $\sum_j x_{j,k} x_{j,k'}$ and $\sum_j g_{i,j}(h_0)^{4/2^p} x_{j,k} x_{j,k'}$, respectively, and the $k$-th elements of the vectors $\mathbf{m}^{(0)}$ and $\mathbf{m}_i^{(p)}$ by $\sum_j x_{j,k} y_j$ and $\sum_j g_{i,j}(h_0)^{4/2^p} x_{j,k} y_j$, respectively.

It is possible to calculate the standard errors of the coefficients, degrees of freedom, and other statistics to obtain the modeling results in linear time. It is also possible to calibrate the parameters by maximizing the corrected Akaike information criterion (AICc). See Appendix 2 and 3 for further details.

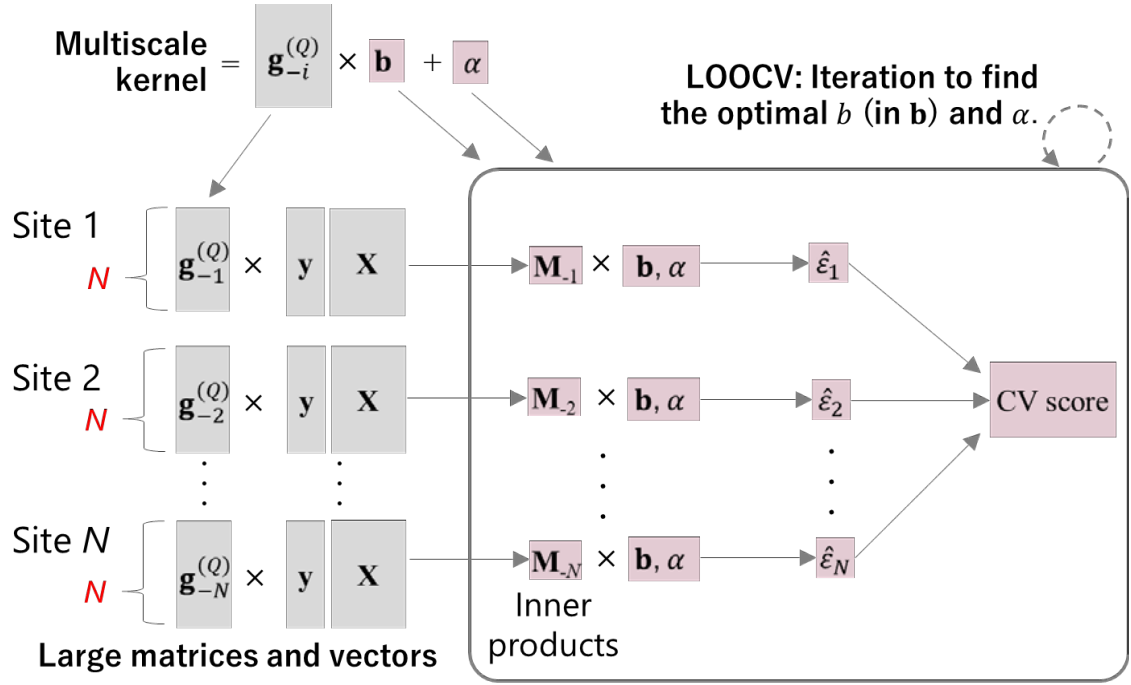

Figure 3. LOOCV routine in ScaGWR. Grey squares represent large matrices and vectors whose sizes depend on *N*, and red squares represent the other small elements. Our multiscale kernel $\mathbf{g}_{-i}^{(Q)}\mathbf{b} + \alpha\mathbf{1}$ includes parameters $\alpha$ and $b$, optimized by the LOOCV. $\mathbf{g}_{-i}^{(Q)}$ is a matrix of *P* polynomials and $\mathbf{b} = [b^1, \cdots, b^P]'$ is the vector consisting of their coefficients.

## Properties of ScaGWR

### *Comparison with a practical alternative*

Basic GWR achieves linear-time LOOCV (or AICc minimization) if the observations other than the *Q*-nearest neighbors are ignored. GWR in the Spatial Statistics Toolbox in ArcGIS (https://www.esri.com/en-us/home) adopts this approach and considers only the 1,000 nearest neighbors by default. The computational complexity for each LOOCV iteration is $O(NK^2Q)$ as it involves performing weighted least-squares estimation *N* times for *Q* sub-samples (each sub-sample has complexity $O(K^2Q)$). Therefore, the complexity depends on *Q*.

In contrast, due to the pre-processing in step (I), which eliminates matrices with dimensions depending on *Q*, a single LOOCV iteration in or algorithm requires only $O(NK^3)$, which is independent of *Q*. Thus, our LOOCV is considerably faster than that of the practical approach. For example, if *K* = 4 and *Q* = 1,000, which is a typical setting, our cost for a single LOOCV iteration is only 0.004 (=$4^3/(4^2 \times 1000)$) times that of the practical GWR.

### *Relationship with geostatistical approaches*

In geostatistics, local and global approaches have been developed for large-scale spatial modeling. ScaGWR is related to local approaches that consider only neighboring observations. Representative local approaches include covariance tapering (Furrer, Genton, and Nychka 2006) and nearest-neighbor Gaussian processes (Datta et al. 2016). Generally, these approaches accurately estimate local spatial variations due to their local modeling scheme. On the other hand, the global geostatistical approach uses a smooth function for spatial process modeling. This approach has been extended for SVCs modeling (see Introduction section). However, this approach tends to underestimate local variations (see Stein 2014); our approach is likely to be more accurate than these global SVC approaches.

For fast CP, when applying a local approach, it is standard to determine the number of neighbors $Q$ being considered (or other parameter specifying local sub-samples) a priori. This is because Furrer, Genton, and Nychka (2006), Datta et al. (2016), and others have suggested that local approaches are highly accurate, irrespective of the value of $Q$. Thus, our pre-determining of $Q$ is reasonable.

However, we still need to verify whether ScaGWR is insensitive to the choice of $Q$. The next section examines this through a simulation study.

## Monte Carlo simulation experiment 1

Below, we perform three Monte Carlo experiments. This section explains the first experiment comparing ScaGWR and GWR in various settings. As a supplement, the second experiment, in the next section, compares the sensitivity of ScaGWR and GWR to the spatial scale of SVCs. This experiment is important as ScaGWR assumes a fixed bandwidth value for the base kernel whereas GWR estimates it. Although these experiments use the LOOCV-based ScaGWR, the third experiment compares the LOOCV-based specification with an AICc-based specification.

### *Outline*

This section compares the estimation accuracies and CP times of GWR and ScaGWR through a Monte Carlo simulation. For ScaGWR, we considered the following cases: number of neighbors $Q \in \{50, 100, 200\}$ and number of polynomials $P \in \{1,2,3,4,6\}$. A Gaussian kernel was used in GWR and the same kernel was employed as the base kernel $g_{i,j}(h_0)$ in ScaGWR. We used a Mac Pro (3.5 GHz, 6-Core Intel Xeon E5 processor with 64 GB of memory) for the CP, R (version 3.6.2; https://cran.r-project.org/) for model estimation, and the R package GWmodel (version 2.1-3; see Lu et al. 2014b; Gollini et al. 2015) for GWR calibration. The code for ScaGWR was embedded into C++ code via the Rcpp package (Eddelbuettel 2013) and implemented in the GWmodel package. Note that we did not parallelize the computation.

The *X* and *Y* coordinates of the sample sites were randomly determined using two standard multivariate normal distributions, $N(\mathbf{0}, \mathbf{I})$. At each site, the true data were generated through Eq. (12):

$$\mathbf{y} = \boldsymbol{\beta}_0 + \mathbf{x}_1 \cdot \boldsymbol{\beta}_1 + \mathbf{x}_2 \cdot \boldsymbol{\beta}_2 + \boldsymbol{\varepsilon} \quad \boldsymbol{\varepsilon} \sim N(\mathbf{0}, \sigma^2 \mathbf{I}),$$
$$\boldsymbol{\beta}_0 \sim N(\mathbf{1}, 0.5^2 \mathbf{G}), \quad \boldsymbol{\beta}_1 \sim N(\mathbf{1}, 2^2 \mathbf{G}), \quad \boldsymbol{\beta}_2 \sim N(\mathbf{1}, 0.5^2 \mathbf{G}), \quad (12)$$

where "$\cdot$" is the element-wise product. $\boldsymbol{\beta}_1$, whose variance is $2^2$, explains more spatial variations than $\boldsymbol{\beta}_0$ and $\boldsymbol{\beta}_2$, whose variances are $0.5^2$. Subsequently, we refer to $\boldsymbol{\beta}_1$ as strong SVCs and to $\{\boldsymbol{\beta}_0, \boldsymbol{\beta}_2\}$ as weak SVCs. The (*i*, *j*)-th element of $\mathbf{G}$ is $g(d_{i,j}) = exp(-d_{i,j}^2)$, suggesting a Gaussian kernel with true bandwidth value of 1.0. ScaGWR was applied for $N \in \{500, 1,000, 3,000, 5,000, 7,000, 10,000, 20,000, 40,000, 60,000, 80,000\}$ whereas GWR was applied in the six cases with $N \leq 10,000$. For each *N*, the estimation included 200 iterations. LOOCV (see Figure 1) was used for bandwidth selection in GWR whereas another LOOCV, illustrated in Figure 2, was used to select *α* and *b*, which are substituted for the bandwidth parameters in ScaGWR.

To calculate the SVCs estimation error, we used root mean squared error (RMSE):

$$RMSE[\hat{\beta}_{i,k}] = \sqrt{\frac{1}{200N} \sum_{iter=1}^{200} \sum_{i=1}^{N} (\hat{\beta}_{i,k}^{iter} - \beta_{i,k})^2}, \qquad (13)$$

where $\hat{\beta}_{i,k}^{iter}$ is the estimated coefficient in the *iter*-th iteration.

## *Results*

Figures 4 and 5 summarize the RMSEs for small to moderate samples ($N \leq 10{,}000$) and larger samples ($80{,}000 \leq N$), respectively. They show that the strong and weak SVCs have different tendencies.

The ScaGWR estimates for the strong SVCs ($\boldsymbol{\beta}_1$) become more accurate as *P* and *Q* become smaller. Small *P* and *Q* reduce the effective number of parameters in the ScaGWR model; a smaller *P* reduces the number of polynomials whereas a smaller *Q* reduces the share of local weights. The result that a parsimonious specification yields more accuracy is reasonable because GWR tends to suffer from local collinearity, which implies a redundancy in the model. As *N* increases, the accuracy of the ScaGWR estimates for the strong SVCs improves and, approximately when $10{,}000 \leq N$, the RMSE values were similar across values of *P* and *Q*.

In contrast to the estimates of the strong SVCs, those of the weak SVCs ($\boldsymbol{\beta}_0, \boldsymbol{\beta}_2$) exhibited better accuracy when *P* and *Q* were large. Large *P* and *Q* values increase the effective number of parameters. This suggests that to identify weak spatially varying signals, a larger number of parameters are needed. Based on Figure 5, *Q* determined the scalability of the estimation accuracy. Specifically, *Q* = 50, 100, and 200 yielded slower, moderate, and faster decays of RMSE values as *N* increased. To estimate weak SVCs accurately from large samples, a large *Q* is desirable. However, *Q* = 100 achieved reasonable accuracy across the different cases (see Figure 5).

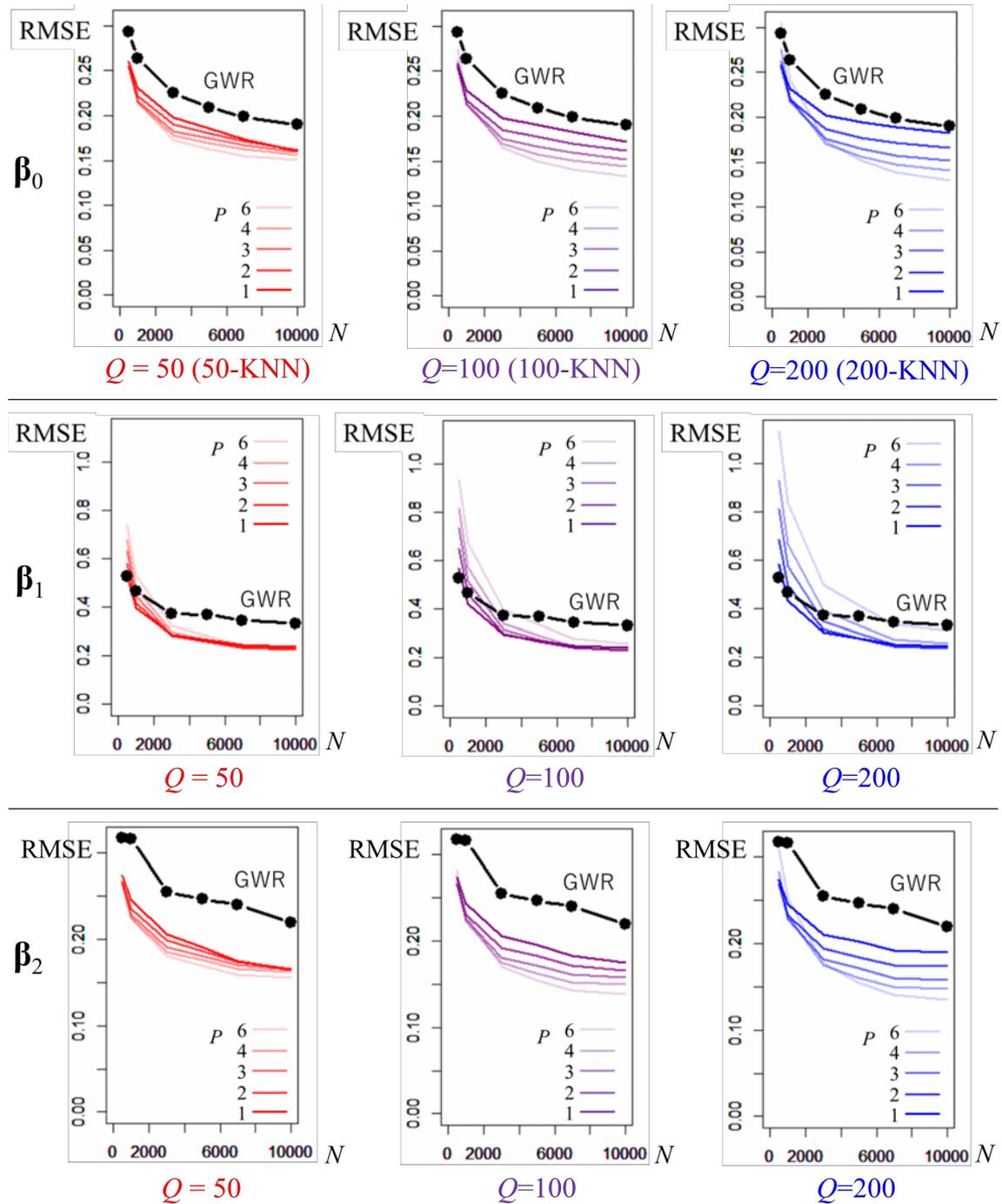

Figure 4. RMSEs of SVCs (10,000 ≤ *N*)

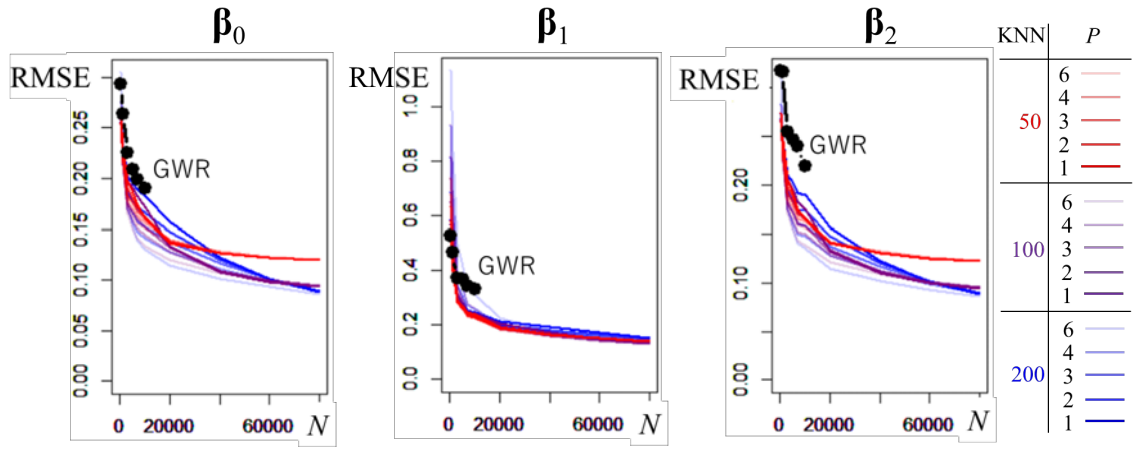

Figure 5. RMSEs of SVCs (80,000 ≤ N).

Among the ScaGWR specifications, we prefer $P = 4$ and $Q = 100$ because it achieves reasonable estimation accuracy regardless of whether $N$ is small or large. Table 1 compares the mean RMSE values from ScaGWR with $P = 4$ and $Q = 100$, with those from GWR. ScaGWR exhibited smaller mean RMSE values in all cases with $N \geq 3{,}000$, confirming the improved accuracy of the ScaGWR. Specifically, for the strong SVCs ($\beta_1$), ScaGWR resulted in smaller RMSEs in all the 200 × 4 experiments with $N \in \{3{,}000, 5{,}000, 7{,}000, 10{,}000\}$. For the weak SVCs ($\beta_2$), ScaGWR exhibited smaller RMSE values than GWR in the 94%, 93%, 91%, and 95% of trials in each 200 experiments in the four cases. ScaGWR's higher accuracy is due to the fact that ScaGWR estimates both the local and global structures using $b$ and $\alpha$ whereas GWR considers only the former. The higher accuracy can also be attributed to our empirical Bayes estimator, which mitigates multicollinearity, whereas the usual GWR estimator is simply a least-squares (non-Bayesian) estimator.

Table 1. Improvement ratios of the RMSEs for ScaGWR with $P = 4$ and $Q = 100$ relative to the RMSEs for GWR ([Mean RMSE value of ScaGWR]/[Mean RMSE value of GWR]).

| $N$ | $\beta_0$ | $\beta_1$ | $\beta_2$ |
|---|---|---|---|
| 500 | 0.89 (0.79) | 1.54 (0.06) | 0.84 (0.92) |
| 1000 | 0.82 (0.93) | 1.24 (0.30) | 0.76 (0.96) |
| 3000 | 0.81 (0.94) | 0.94 (1.00) | 0.72 (0.94) |
| 5000 | 0.79 (0.81) | 0.79 (1.00) | 0.68 (0.93) |
| 7000 | 0.72 (0.69) | 0.72 (1.00) | 0.65 (0.91) |
| 10000 | 0.69 (0.71) | 0.71 (1.00) | 0.60 (0.94) |

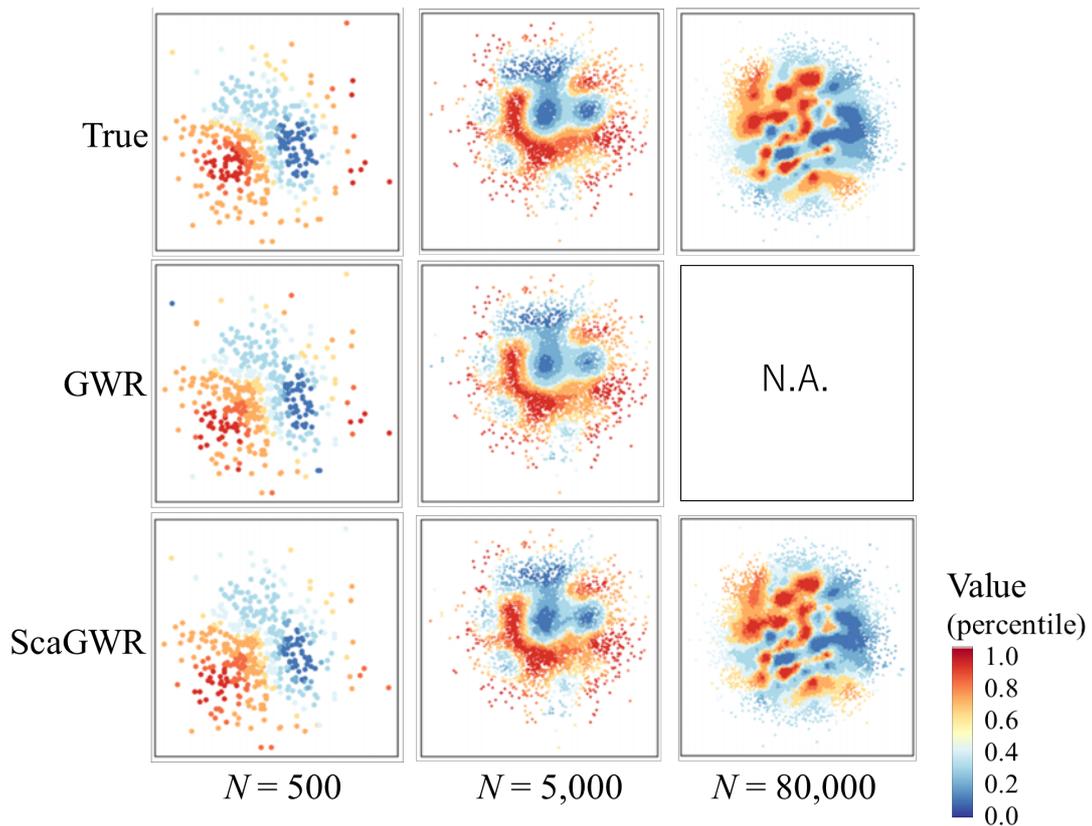

Figure 6. True and estimated SVCs ($\beta_1$) in each first iteration. $P = 4$ and $Q = 100$ were used for ScaGWR.

Figure 6 plots the true and estimated SVCs in the first iterations when $N \in \{500, 5{,}000, 80{,}000\}$. This figure confirms that ScaGWR with $P = 4$ and $Q = 100$ accurately estimates the SVCs. ScaGWR successfully captures small-scale variations even if $N = 80{,}000$ whereas GWR is not feasible because of the large sample size.

Figure 7 shows the standard deviations (SDs) of the estimated SVCs. In Eq. (12), we used $\{0.5, 2.0, 0.5\}$ as the true SDs for $\{\boldsymbol{\beta}_0, \boldsymbol{\beta}_1, \boldsymbol{\beta}_2\}$, respectively. GWR tends to underestimate the SDs for strong SVCs while overestimating the SDs for weak SVCs. Unfortunately, the latter upward bias does not decrease even if $N$ increases to 10,000. On the other hand, ScaGWR tends to underestimate the SDs for all the SVCs when $N$ is small. This is because our empirical Bayes estimator strongly shrinks the estimates toward the OLS estimator for small samples. However, this bias rapidly decreases as $N$ increases. This result demonstrates that ScaGWR successfully estimates the different amounts of variation of SVCs in large samples.

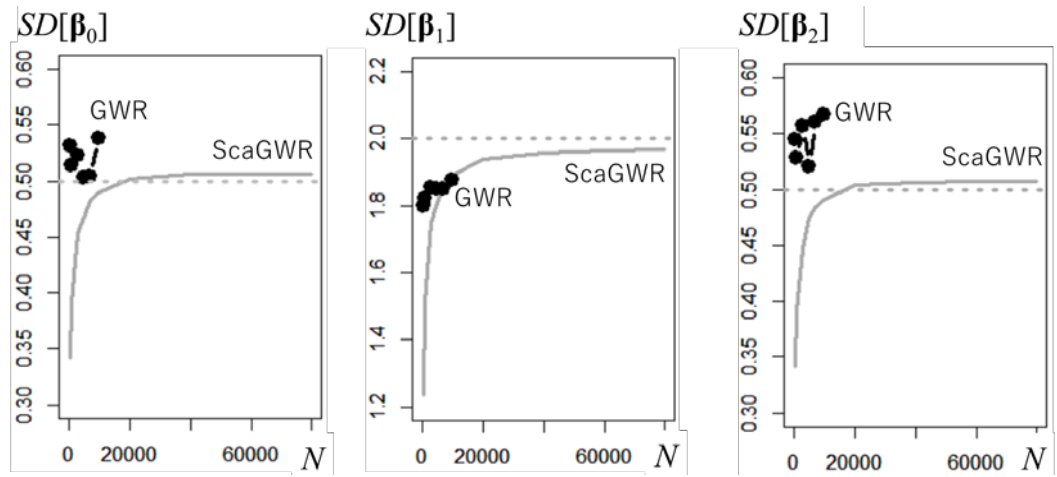

Figure 7. Standard deviations of estimated SVCs. $P = 4$ and $Q = 100$ were used for ScaGWR. The dashed line denotes the true values.

Figure 8 compares the CV scores, which are a measure of the out-of-sample prediction accuracy. This figure suggests that ScaGWR is faster but as accurate as standard GWR in terms of spatial prediction.

Finally, Figure 9 summarizes the average CP times. The CP time of standard GWR rapidly increased with $N$. It took 240.2 seconds when $N = 10,000$. By contrast, the CP time of ScaGWR increased only linearly with $N$. On average, ScaGWR with $Q = 100$ and $P = 4$ took 6.6 seconds for execution when $N = 10,000$, and 89.8 seconds when $N = 80,000$. Even when $N = 1,000,000$, ScaGWR took only 1,637 seconds on average in five trials without parallelization.

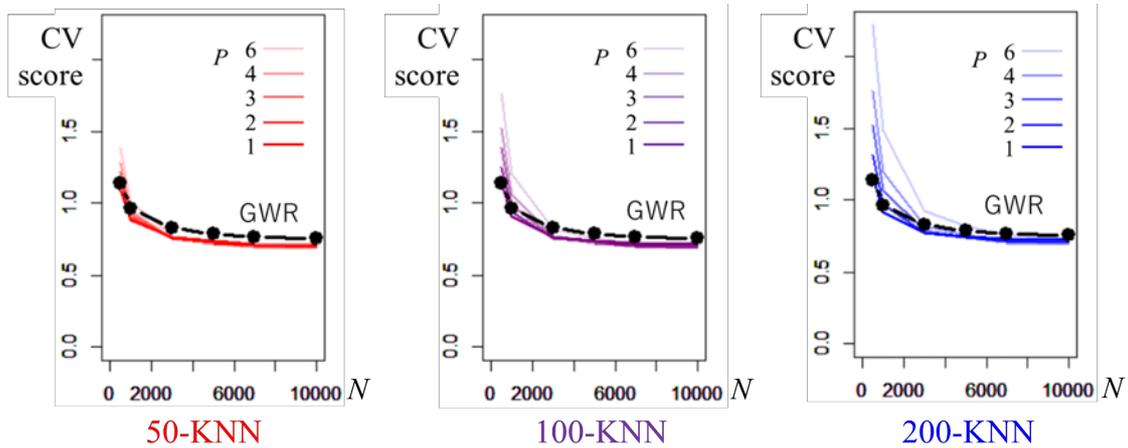

Figure 8. CV scores of GWR (black line) and ScaGWR (colored lines).

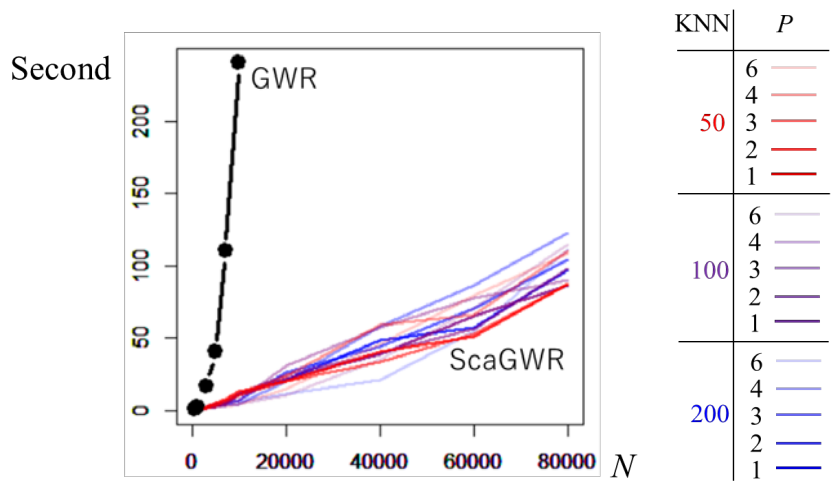

Figure 9. Computation times of LOOCV-based GWR and ScaGWR calibrations.

In summary, ScaGWR with $Q = 100$ and $P = 4$ is faster than basic GWR. In addition, ScaGWR tends to have a higher coefficient estimation accuracy, especially when $N \geq 3{,}000$.

## Monte Carlo experiment 2

### *Outline*

Although we have shown that ScaGWR accurately estimates the coefficients when $Q = 100$, this might not hold if the true SVCs have larger- or smaller-scale spatial patterns. To investigate this, we performed another experiment using 2.0 and 0.5 as the bandwidth values for the true SVCs (see Figure 10 for the map patterns). Recall that, in the previous experiment, we had set both values to 1.0. As same as the previous experiment, we used $Q = 100$ and $P = 4$. The other simulation settings were the same as those in the previous section.

### *Results*

Figure 10 shows a plot of the GWR and ScaGWR estimates in the first iteration when $N$=5,000, together with the true SVCs. This plot demonstrates that GWR and ScaGWR accurately estimate SVCs, irrespective of the true scale. Therefore, our scale parameter $b$ successfully acts as an alternative to the usual bandwidth parameter. Figure 11 shows a comparison of the mean RMSEs from the simulations. Both GWR and ScaGWR improve their estimation accuracy as the scale of the true SVCs increases because large-scale SVCs have simpler map patterns and are easier to estimate. Based on the results, ScaGWR's estimation accuracy is as robust as that of classical GWR to the scale of SVCs.

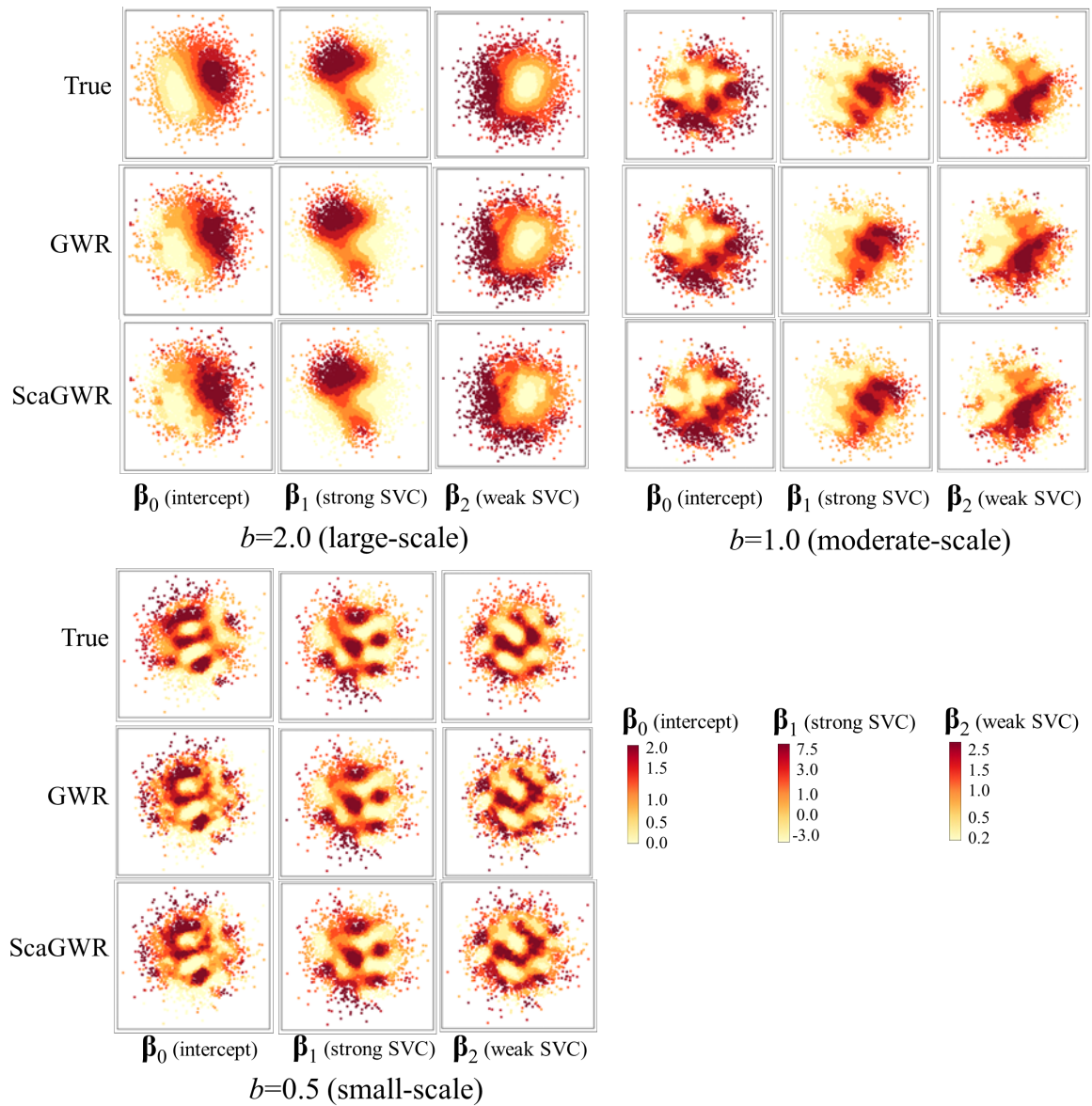

Figure 10. Examples of true and estimated SVCs ($N = 5,000$) when the true bandwidths are 2.0, 1.0, and 0.5. $P = 4$ and $Q = 100$ were used for ScaGWR.

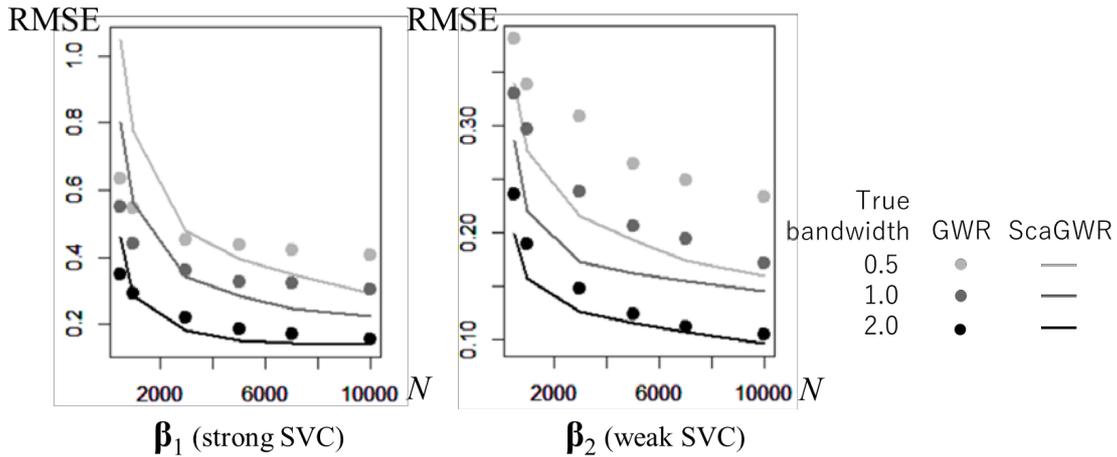

Figure 11. RMSEs of SVCs ($N$ = 5,000 and $b \in \{0.5, 1.0, 2.0\}$).

## Monte Carlo experiment 3

### *Outline*

While Sections 4 and 5 describe the LOOCV-based ScaGWR, this section presents the performance of the AICc-based ScaGWR. The other simulation settings were the same as those of Monte Carlo experiment 2.

### *Results*

We first compared the estimation accuracy of ScaGWR with that of the AICc-based GWR for cases with $N \leq 10,000$. Table 2 summarizes the resulting AICc values. ScaGWR has a smaller AICc when $N \geq 3,000$, which confirms ScaGWR's higher accuracy for large samples. Table 3, which compares the RMSEs of the estimated coefficients, shows that ScaGWR's estimation accuracy exceeds that of classical GWR in four out of the six cases with $N < 3,000$ and for all the cases with $N \geq 3,000$. Figure 12 shows the comparison between the RMSEs of the $\beta_1$s estimated using the AICc-based ScaGWR and the LOOCV-based ScaGWR. This figure demonstrates that the accuracies of these two methods are quite similar.

Table 2. Comparison of the mean AICc values.

| Sample size | GWR | ScaGWR ($Q = 100$, $P = 4$) |
|---:|---:|---:|
| 500 | 1411 | 1555 |
| 1,000 | 2686 | 2783 |
| 3,000 | 7655 | 7616 |
| 5,000 | 12601 | 12485 |
| 7,000 | 17583 | 17245 |
| 10,000 | 25066 | 24487 |

Table 3. Comparison of the mean RMSEs between AICc-based GWR and ScaGWR.

| Sample size | $\beta_0$ | | $\beta_1$ | | $\beta_2$ | |
|---:|---:|---:|---:|---:|---:|---:|
| | GWR | ScaGWR | GWR | ScaGWR | GWR | ScaGWR |
| 500 | 0.35 | 0.26 | 0.69 | 0.81 | 0.37 | 0.28 |
| 1,000 | 0.28 | 0.22 | 0.45 | 0.56 | 0.33 | 0.22 |
| 3,000 | 0.26 | 0.16 | 0.39 | 0.35 | 0.40 | 0.17 |
| 5,000 | 0.28 | 0.15 | 0.42 | 0.30 | 0.32 | 0.16 |
| 7,000 | 0.21 | 0.14 | 0.35 | 0.26 | 0.26 | 0.15 |
| 10,000 | 0.21 | 0.14 | 0.39 | 0.24 | 0.32 | 0.14 |

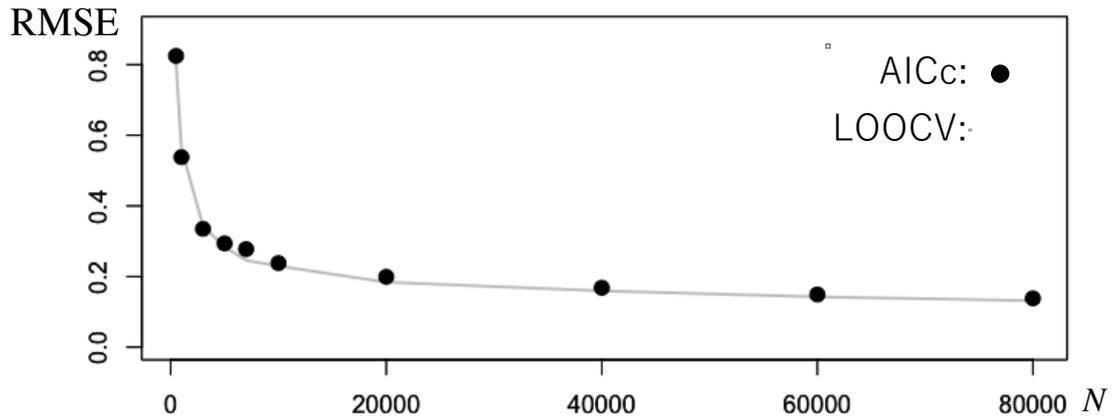

Figure 12. RMSE of **β**$_1$ for LOOCV-based and AICc-based ScaGWR (*N* = 5,000).

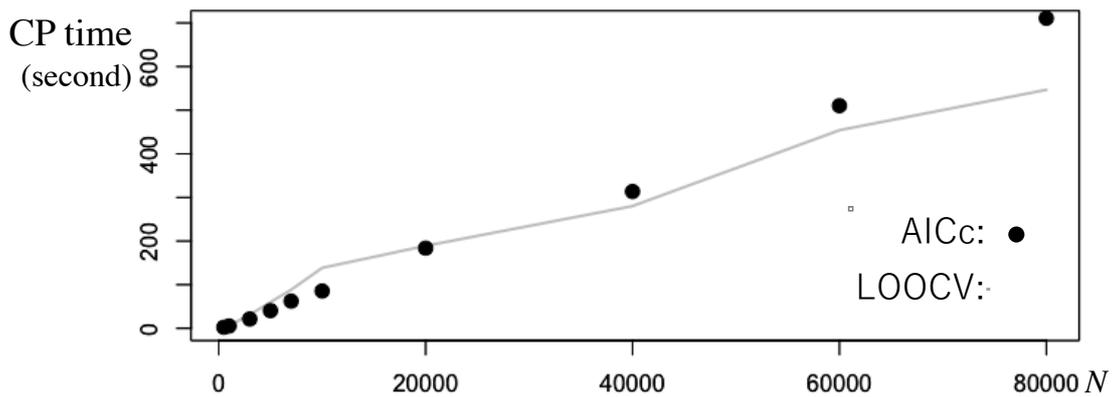

Figure 13. Computation times (LOOCV-based versus AICc-based ScaGWR).

Finally, the computational times of the two ScaGWR specifications are compared in Figure 13. Our AICc minimization successfully achieves a linear-time estimation. However, the computational time is slightly longer than that for the LOOCV when $N \geq 40{,}000$ because of the additional calculations, as explained in Appendix 3. Based on these results, LOOCV might be a better choice for very large samples whereas the AICc-based approach is suitable for likelihood-based inference.

## Empirical analysis

### *Outline*

This section applies the LOOCV-based ScaGWR to a 2015 US household income dataset. Based on our simulation results above, we set $Q = 100$ and $P = 4$. A Gaussian kernel was also used. The dependent variable is tract-level census median household income (Figure 14). The sample size was 72,359. Classical GWR is not feasible for these data because of the computational burden. The covariates were as follows: (1) Univ: Ratio of people with a Bachelor's degree or higher educational attainment among people over age 25. (2) Eng: Ratio of people who speak English at home to the total population. (3) Age: Average age. The SVCs for these covariates are denoted by $\{\beta_{Univ}, \beta_{Eng}, \beta_{Age}\}$ whereas the spatially varying intercept is denoted by $\beta_0$.

We expect the influence of Univ to be positive. This is because people who are university graduates have a greater chance of obtaining white-collar or professional jobs. Eng is also expected to have a positive impact since higher English skills allow people to communicate and interact with a wider variety of business partners. Regarding Age, we expect its influence to become either positive or negative depending on the region. This is because older workers tend to have higher earnings due to their longer-term working experience whereas younger workers tend to have greater earnings because they are more familiar with recent technologies and facilities/devices and knowledge of these can increase earnings. Given the considerable heterogeneity of income across the United States, the influences of these three variables are likely to vary across space. Therefore, ScaGWR is suitable for this analysis.

All the data were obtained from the 2015 US census using the tidycensus package (https://cran.r-project.org/web/packages/tidycensus/index.html).

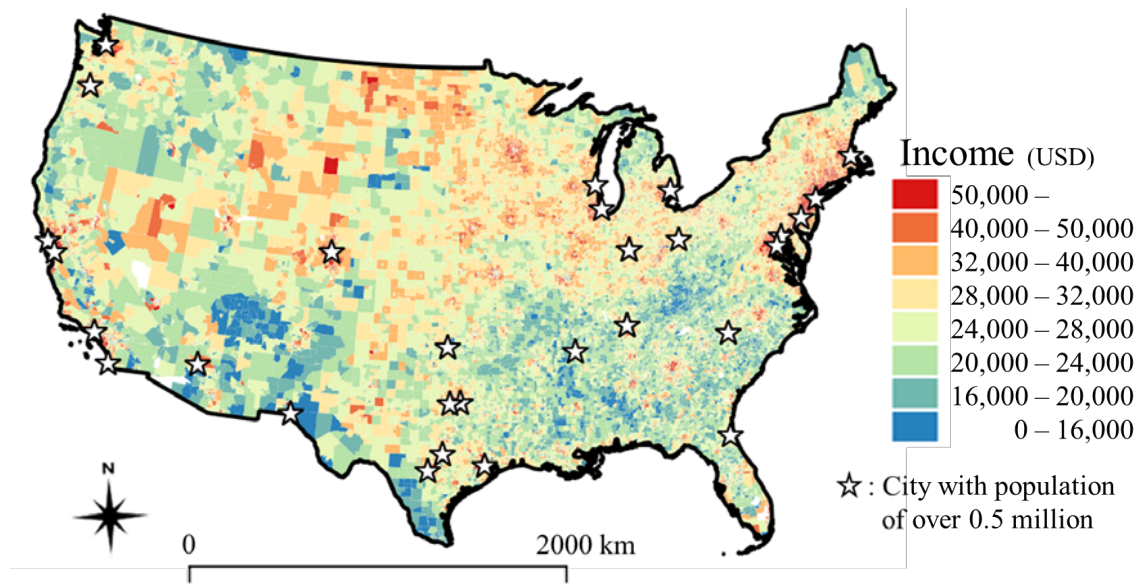

Figure 14. Tract-level median household income in 2015.

*Results*

ScaGWR takes 20.6 seconds on the PC mentioned in Section 4. Table 4 compares the ScaGWR estimates with those from OLS. With OLS, Univ and Age have a statistically significant positive impact whereas Eng is statistically nonsignificant. The ScaGWR estimates suggest considerable spatial variation in the regression coefficients whereas the medians are similar to their OLS estimates. Both the adjusted $R^2$ values and the AICc values are improved when considering the aforementioned spatial variation. The ScaGWR estimates are more reliable.

Table 4. Comparison of the coefficients estimated from OLS and ScaGWR.

|  | OLS | | ScaGWR | | | | | |
| --- | --- | --- | --- | --- | --- | --- | --- | --- |
|  | Est. | t-value | Min. | 1st | Median | Mean | 3rd | Max. |
| Intercept | 7,931 | 37.6 ***[1)] | -99,750 | -6,379 | 6,297 | 2,682 | 15,042 | 49,810 |
| Univ | 48,963 | 333.9 *** | -11,223 | 33,237 | 41,902 | 42,493 | 50,486 | 142,370 |
| Eng | 204 | 1.4 | -64,610 | -2,338 | 4,941 | 5,553 | 12,719 | 76,826 |
| Age | 189 | 33.4 *** | -1,618 | -80 | 185 | 251 | 526 | 2,907 |
| Adj. $R^2$ | 0.626 | | 0.802 | | | | | |
| AICc | 1,487,283 | | 1,444,063 | | | | | |

[1)]*** denotes statistical significance of 1 %

Figure 15 summarizes the estimated SVCs. Here, colors are used only for census tracts whose coefficients are statistically significant at the 5 % level. Following Nakaya's (2007) suggestion that spatially varying intercepts should not be interpreted because the estimates are highly dependent on the local correlations among the explanatory variables, we do not display the estimated $\beta_0$ here. $\beta_{Univ}$ indicates a large positive impact near cities with populations exceeding 0.5 million, suggesting that highly educated people living in large cities earn higher incomes. Given the concentration of white-collar and professional workers in large cities, this result is reasonable. $\beta_{Eng}$ also indicates large positive values around major cities. Because speaking English is important for white-collar workers to extend their jobs through national or international contacts, this result is also reasonable. $\beta_{Eng}$ also suggests that speaking English increases earnings on the north side of the United States. By contrast, the coefficient becomes small or negative on the south side of the United States. Based on the low income levels in the south (see Figure 14), this result suggests that residents in the southeast area tend to have lower incomes irrespective of whether they speak English. Although $\beta_{Age}$ had a positive impact in the eastern United States, its value tends to be small near major cities on a local scale. Based on this result, older workers, who tend to be long-term or veteran workers, have higher earnings in eastern suburban areas.

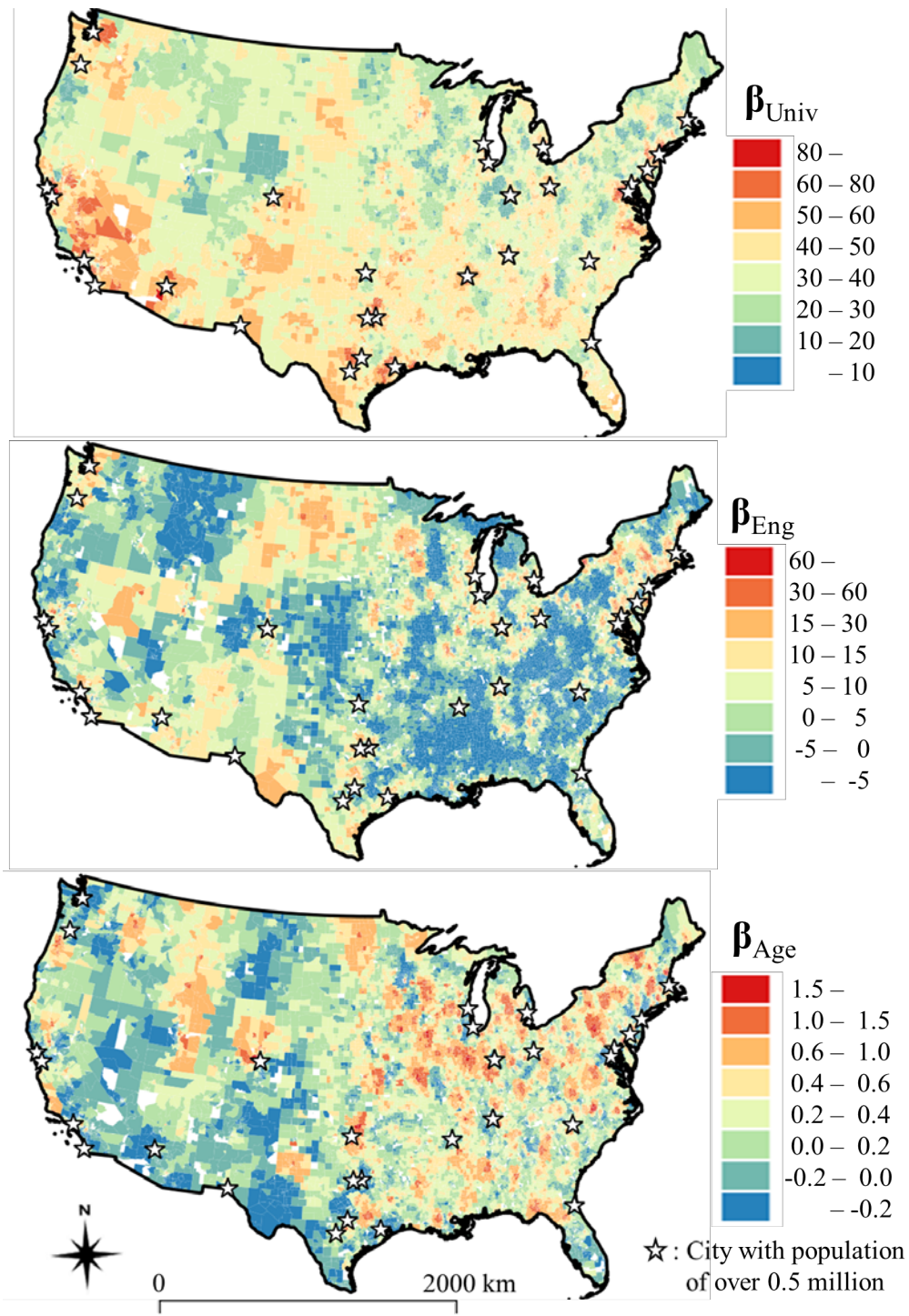

Figure 15. Estimated SVCs ($Q = 100$; LOOCV). Colors are drawn only on census tracts whose coefficients are statistically significant at the 5 % level.

AICc-based ScaGWR was also applied to the data. The spatial distribution of $\boldsymbol{\beta}_{Univ}$, which is plotted in Figure 16, exhibits almost the same pattern as the one in the LOOCV-based results. This suggests robustness of the results to the calibration approach used. Finally, LOOCV-based ScaGWR with $Q = 500$ was applied, and the estimated $\boldsymbol{\beta}_{Univ}$ is plotted in Figure 17. The plot confirms that the spatial pattern of the estimates becomes similar even if the $Q$ value is changed.

In summary, this section empirically verifies that ScaGWR successfully captures local-scale spatial patterns in the regression coefficients. Recall that other fast SVCs modeling approaches relying on a smooth function for modeling SVCs tend to fail to capture such local variations (see the introduction section). Furthermore, ScaGWR's CP time was quite short whereas classical GWR was not feasible for this large dataset without sampling, aggregation, or another effort to reduce the sample size. ScaGWR, which was fast and accurate, can be useful for a wide variety of large-scale spatial modeling.

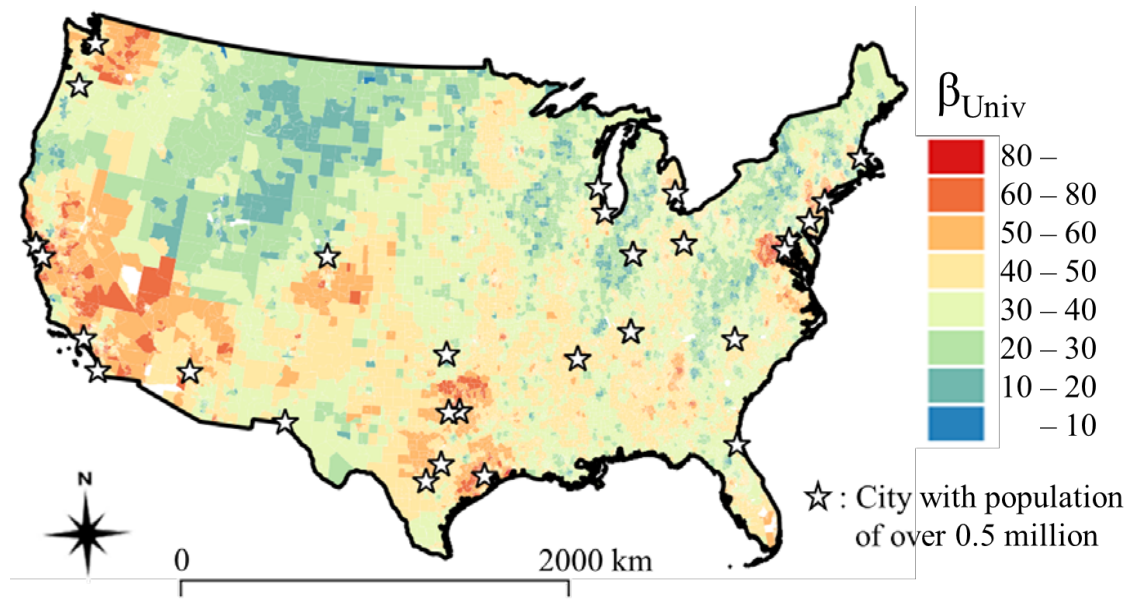

Figure 16. Estimated SVCs ($Q = 500$; LOOCV). Colors are drawn only on census tracts whose coefficients are statistically significant at the 5 % level.

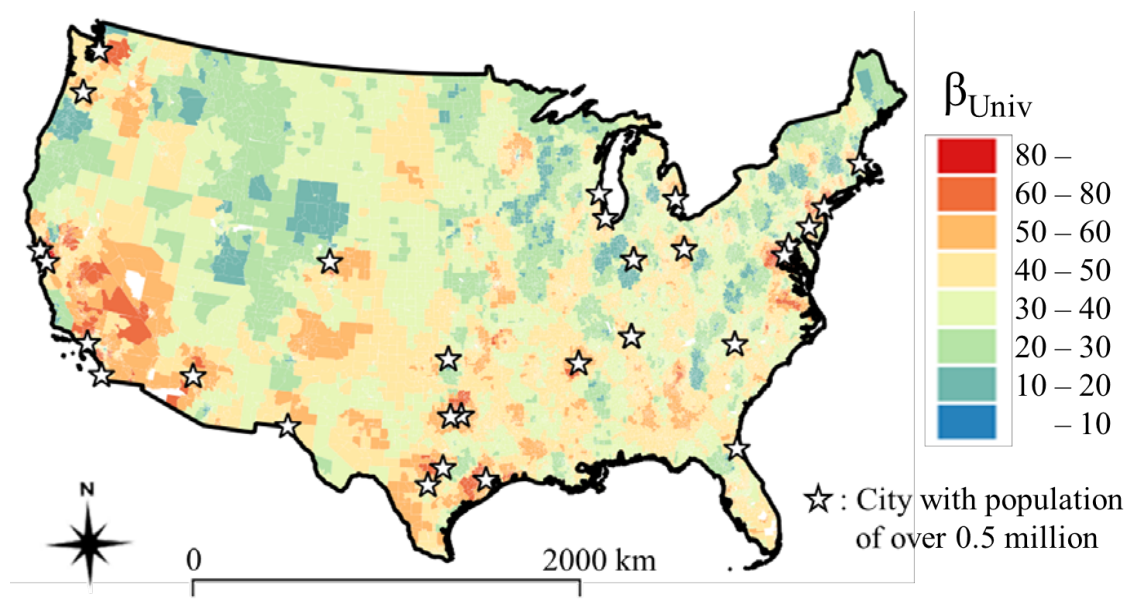

Figure 17. Estimated SVCs ($Q = 100$; AICc). Colors are drawn only on census tracts whose coefficients are statistically significant at the 5 % level.

## Concluding remarks

As the size of spatial data becomes increasingly larger, computational efficiency also becomes increasingly important in spatial modeling. Because large samples imply a greater chance of capturing heterogeneity over geographical space, the assumption of spatial stationarity, which is a standard assumption in geostatistics, is appropriate for modeling large spatial datasets.

Given this background, this study proposed ScaGWR. Unlike existing GWR algorithms, ScaGWR achieves quasi-linear computational complexity, as illustrated in Figure 9. This property allows us to estimate SVCs from millions of observations, even without parallelization.

Most fast alternatives, which model SVCs using smooth functions, cannot capture small-scale map patterns (this is the degeneracy problem; see Stein 2014). By contrast, our local modeling approach does not suffer from this problem, as confirmed by the results in Figure 6. Thus, ScaGWR satisfies the requirements listed in the introduction section, namely, (i) a linear-time computational burden and (ii) avoidance of the degeneracy problem.

In practice, ScaGWR is superior for large samples (e.g., $N \geq 3{,}000$). Based on our simulation study, ScaGWR with $Q = 100$ and $P = 4$ should be the default choice if $N \leq 100{,}000$. This choice of $Q$ and $P$ will be sensible for even larger values of $N$ for the following reasons: these values are sufficiently small to handle larger samples because these parameters have direct impacts only on the $Q$-nearest neighbors. Therefore, if $Q \ll N$, the performance of ScaGWR will be similar even if $N$ is very large.

A drawback of standard GWR is its tendency to suffer from local multicollinearity. Even if a regularization parameter is imposed, GWR, which considers only samples within a kernel window, might be unstable if the number of observations within the window is small. By contrast, ScaGWR implicitly imposes $\boldsymbol{\beta}^{prior} \sim N(\widehat{\boldsymbol{\beta}}_{OLS}, \alpha^{-1}(\mathbf{X}'\mathbf{X})^{-1})$ as a prior (see the "Coefficients estimator" section). Therefore, even if the kernel window is small (i.e., $b$ is small), the estimator, which shrinks to the OLS estimator, is likely to be stable. The robustness of ScaGWR needs to be studied in future work. Another important topic is the absence of the bandwidth parameter, which is useful for interpreting the estimated SVCs. The development of a spatial scale measure for ScaGWR will be useful for large-scale exploratory analyses.

Comparing ScaGWR with fast alternatives using very large samples (e.g., in the order of millions of observations) is an important future topic. The GWR based on the 1,000 nearest neighbors, implemented in the Spatial Statistics Toolbox in ArcGIS, and the fast GWR implemented in mgwr (Oshan et al. 2019a), a Python module in PySAL (Rey 2019), are also available for large-scale GWR modeling.

Extending ScaGWR is also an interesting subject for future research. Its computational efficiency might be preserved even if it is extended to multiscale GWR (Fotheringham, Yang, and Kang 2017; Lu et al. 2019; Oshan et al. 2019b), such as GWR with parameter-specific distance matrices (Lu et al. 2017, 2018), GWR with a flexible bandwidth (Yang 2014; Murakami et al. 2019), and conditional GWR (Leong and Yue 2017). Furthermore, ScaGWR might be important in combining GWR with a global model (see Geniaux and Martinetti 2018; Harris 2019). The ScaGWR model can be considered an integration of the GWR model and a global linear regression model. The linear regression model can be potentially replaced with a mixed effects model, additive model, temporal model, and so on. Such multi-model integration is an interesting development toward a fast, stable, and flexible reformulation of GWR.

ScaGWR was implemented in the R package **scgwr** (https://cran.r-project.org/web/packages/scgwr/). We also implemented ScaGWR, which was embedded into C++ code via the Rcpp package, in the **GWmodel** package (https://cran.r-project.org/web/packages/GWmodel/).

# Acknowledgement

This work was supported by the Japan Society for the Promotion of Science under Grant 17H02046, 18H01556, and 18H03628; the Joint Support Center for Data Science Research under Grant 006RP2018 and 004RP2019.

## Appendix 1. Derivation of Eq. (10)

This appendix shows the derivation of Eq. (10) by expanding $\mathbf{X}'\mathbf{G}_{-i}(b,\alpha)\mathbf{X}$ and $\mathbf{X}'\mathbf{G}_{-i}(b,\alpha)\mathbf{y}$ from Eq. (9). The ($k$, $k'$)-th element of $\mathbf{X}'\mathbf{G}_{-i}(b,\alpha)\mathbf{X}$ is analytically obtained as

$$\sum_{j \neq i} g_{i,j}(b,\alpha) x_{j,k} x_{j,k'} \qquad (A1)$$

By substituting Eq. (6) into Eq. (10), it is further expanded as

$$\sum_{j \neq i} \sum_{p=1}^{P} \left( \alpha + b^p g_{i,j}^{(Q)}(h_0)^{4/2^p} \right) x_{j,k} x_{j,k'}$$

$$= \alpha \sum_{j \neq i} x_{j,k} x_{j,k'} + \sum_{p=1}^{P} b^p \sum_{j \neq i} g_{i,j}(h_0)^{4/2^p} x_{j,k} x_{j,k'}, \qquad (A2)$$

where $m_{-i,k,k'}^{(0)} = \sum_{j \neq i} x_{j,k} x_{j,k'}$ and $m_{-i,k,k'}^{(p)} = \sum_{j \neq i} g_{i,j}(h_0)^{4/2^p} x_{j,k} x_{j,k'}$. Similarly, the $k$-th element of $\mathbf{X}'\mathbf{G}_i(b,\alpha)\mathbf{y}$ is given by

$$\sum_{j \neq i} g_{i,j}(b,\alpha) x_{j,k} y_j = \alpha m_{-i,k,\mathbf{y}}^{(0)} + \sum_{p=1}^{P} b^p m_{-i,k,\mathbf{y}}^{(p)}, \qquad (A3)$$

where $m_{-i,k,\mathbf{y}}^{(0)} = \sum_{j \neq i} x_{j,k} y_j$ and $m_{-i,k,\mathbf{y}}^{(p)} = \sum_{j \neq i} g_{i,j}(h_0)^{4/2^p} x_{j,k} y_j$. By substituting Eqs. (11) and (12) into Eq. (9), the expression for $\widehat{\boldsymbol{\beta}}_{-i}$ becomes:

$$\widehat{\boldsymbol{\beta}}_{-i} = \begin{bmatrix} \alpha m_{-i,1,1}^{(0)} + \sum_{p=1}^{P} b^p m_{-i,1,1}^{(p)} & \cdots & \alpha m_{-i,1,K}^{(0)} + \sum_{p=1}^{P} b^p m_{-i,1,K}^{(p)} \\ \vdots & \ddots & \vdots \\ \alpha m_{-i,K,1}^{(0)} + \sum_{p=1}^{P} b^p m_{-i,K,1}^{(p)} & \cdots & \alpha m_{i-,K,K}^{(0)} + \sum_{p=1}^{P} b^p m_{-i,K,K}^{(p)} \end{bmatrix}^{-1}$$

$$\begin{bmatrix} \alpha m_{-i,1,\mathbf{y}}^{(0)} + \sum_{p=1}^{P} b^p m_{-i,1,\mathbf{y}}^{(p)} \\ \vdots \\ \alpha m_{-i,K,\mathbf{y}}^{(0)} + \sum_{p=1}^{P} b^p m_{-i,K,\mathbf{y}}^{(p)} \end{bmatrix}. \qquad (A4)$$

Eq. (A4) is identical to Eq. (10).

## Appendix 2. Diagnostic statistics for ScaGWR

Because the ScaGWR model is equivalent to basic GWR with a multiscale kernel, its diagnostics are the same as well. The effective sample size for the ScaGWR model, which is required to evaluate the standard errors of the coefficients, is defined as in WR as

$$N^* = N - 2tr[\mathbf{S}] + tr[\mathbf{S}'\mathbf{S}], \tag{A5}$$

where $\mathbf{S}$ is the design matrix of the ScaGWR model, specified as

$$\mathbf{S} = \begin{bmatrix} s_{1,1} & s_{1,2} & \cdots & s_{1,N} \\ s_{2,1} & s_{2,2} & \cdots & s_{2,N} \\ \vdots & \vdots & \ddots & \vdots \\ s_{N,1} & s_{N,2} & \cdots & s_{N,N} \end{bmatrix} = \begin{bmatrix} \mathbf{s}'_1 \\ \mathbf{s}'_2 \\ \vdots \\ \mathbf{s}'_N \end{bmatrix} = \begin{bmatrix} \mathbf{x}_1(\mathbf{X}'\mathbf{G}_1(b,\alpha)\mathbf{X})^{-1}\mathbf{X}'\mathbf{G}_1(b,\alpha) \\ \mathbf{x}_2(\mathbf{X}'\mathbf{G}_2(b,\alpha)\mathbf{X})^{-1}\mathbf{X}'\mathbf{G}_2(b,\alpha) \\ \vdots \\ \mathbf{x}_N(\mathbf{X}'\mathbf{G}_N(b,\alpha)\mathbf{X})^{-1}\mathbf{X}'\mathbf{G}_N(b,\alpha) \end{bmatrix}. \tag{A6}$$

$\mathbf{S}$ is a large matrix ($N \times N$) that cannot be stored for large values of *N*. However, $tr[\mathbf{S}]$ can be calculated without explicitly processing $\mathbf{S}$ as follows:

$$\begin{aligned} tr[\mathbf{S}] &= \sum_i s_{i,i} = \sum_i \mathbf{x}_i(\mathbf{X}'\mathbf{G}_i(b,\alpha)\mathbf{X})^{-1}\mathbf{x}'_i g_{i,i}(b,\alpha) \\ &= \sum_i \mathbf{x}_i \left[ \alpha \mathbf{M}^{(0)} + \sum_{p=1}^P b^p \mathbf{M}_i^{(p)} \right]^{-1} \mathbf{x}'_i g_{i,i}(b,\alpha), \end{aligned} \tag{A7}$$

where $\left[ \alpha \mathbf{M}^{(0)} + \sum_{p=1}^P b^p \mathbf{M}_i^{(p)} \right]^{-1}$ is a small matrix ($K \times K$) that has already been calculated when estimating $\widehat{\boldsymbol{\beta}}_i$. Thus, through Eq. (A7), calculating $tr[\mathbf{S}]$ is computationally efficient.

Likewise, $tr[\mathbf{S}'\mathbf{S}]$ can be calculated without processing the large matrix $\mathbf{S}'\mathbf{S}$, by using $tr[\mathbf{S}'\mathbf{S}] = \sum_i \sum_j s_{i,j}^2$ as follows:

$$tr[\mathbf{S}'\mathbf{S}] = \sum_i \sum_j s_{i,j}^2 = \sum_i \mathbf{s}'_i \mathbf{s}_i$$

$$= \sum_i \mathbf{x}_i \left[ \alpha \mathbf{M}^{(0)} + \sum_{p=1}^P b^p \mathbf{M}_i^{(p)} \right]^{-1} \mathbf{X}' \mathbf{G}_i(b,\alpha)^2 \mathbf{X} \left[ \alpha \mathbf{M}^{(0)} \right.$$

$$\left. + \sum_{p=1}^P b^p \mathbf{M}_i^{(p)} \right]^{-1} \mathbf{x}'_i$$

$$= \sum_i \mathbf{x}_i \left[ \alpha \mathbf{M}^{(0)} + \sum_{p=1}^P b^p \mathbf{M}_i^{(p)} \right]^{-1} \mathbf{X}'(\alpha \mathbf{I}$$

$$+ \mathbf{G}_i^{(Q)}(b))^2 \mathbf{X} \left[ \alpha \mathbf{M}^{(0)} + \sum_{p=1}^P b^p \mathbf{M}_i^{(p)} \right]^{-1} \mathbf{x}'_i \qquad (A8)$$

$$= \sum_i \mathbf{x}_i \left[ \alpha \mathbf{M}^{(0)} + \sum_{p=1}^P b^p \mathbf{M}_i^{(p)} \right]^{-1} (\alpha^2 \mathbf{X}'\mathbf{X} + 2\alpha \mathbf{X}' \mathbf{G}_i^{(Q)}(b) \mathbf{X}$$

$$+ \mathbf{X}' \mathbf{G}_i^{(Q)}(b)^2 \mathbf{X}) \left[ \alpha \mathbf{M}^{(0)} + \sum_{p=1}^P b^p \mathbf{M}_i^{(p)} \right]^{-1} \mathbf{x}'_i$$

$$= \sum_i \mathbf{x}_i \left[ \alpha \mathbf{M}^{(0)} + \sum_{p=1}^P b^p \mathbf{M}_i^{(p)} \right]^{-1} (\alpha^2 \mathbf{M}^{(0)} + 2\alpha \sum_{p=1}^P b^p \mathbf{M}_i^{(p)}$$

$$+ \sum_{p=1}^P b^{2p} \mathbf{M}_i^{(2p)}) \left[ \alpha \mathbf{M}^{(0)} + \sum_{p=1}^P b^p \mathbf{M}_i^{(p)} \right]^{-1} \mathbf{x}'_i.$$

Although we also need to calculate $\mathbf{M}_i^{(2p)}$, a $K \times K$ matrix whose $(k, k')$-th element is $\sum_j g_{i,j}(h_0)^{8/2^p} x_{j,k} x_{j,k'}$, the computational complexity is identical to that of $\mathbf{M}_i^{(p)}$. Thus, the calculation is still trivial.

Given $N^*$, the unbiased estimate of the residual variance is

$$\hat{\sigma}^2 = \frac{1}{N^*}\sum_{k=1}^{K}(y_i - \mathbf{x}_i\hat{\boldsymbol{\beta}}_i)^2, \tag{A9}$$

which is equivalent to Eq. (13). The variance estimates for the SVCs are derived as

$$Var[\hat{\boldsymbol{\beta}}_i] = \hat{\sigma}^2(\mathbf{X}'\mathbf{G}_i(b,\alpha)\mathbf{X})^{-1}\mathbf{X}'\mathbf{G}_i(b,\alpha)^2\mathbf{X}(\mathbf{X}'\mathbf{G}_i(b,\alpha)\mathbf{X})^{-1},$$

$$= \hat{\sigma}^2\left[\alpha\mathbf{M}^{(0)} + \sum_{p=1}^{P}b^p\mathbf{M}_i^{(p)}\right]^{-1}\left(\alpha^2\mathbf{M}^{(0)} + 2\alpha\sum_{p=1}^{P}b^p\mathbf{M}_i^{(p)}\right. \tag{A10}$$

$$\left. + \sum_{p=1}^{P}b^{2p}\mathbf{M}_i^{(2p)}\right)\left[\alpha\mathbf{M}^{(0)} + \sum_{p=1}^{P}b^p\mathbf{M}_i^{(p)}\right]^{-1},$$

which is again trivial to calculate. Thus, diagnostics for ScaGWR are available for large samples. Note that the CP time presented in Section 4 includes the time taken to estimate both $\hat{\boldsymbol{\beta}}_i$ and $Var[\hat{\boldsymbol{\beta}}_i]$.

## Appendix.3. AICc-based approach to optimize $\{b, \alpha\}$

The parameters $\{b, \alpha\}$ can also be estimated using AICc, which is widely accepted for bandwidth estimation in GWR (e.g., Nakaya et al. 2005). The AICc-based procedure consists of the same steps, (I), (II), and (III), except that the second step is replaced with the minimization of the AICc, whose formula is

$$\text{AICc} = Nlog(\hat{\sigma}^2) + Nlog(2\pi) + N\frac{N + tr[\mathbf{S}]}{N - 2 - tr[\mathbf{S}]}. \tag{A11}$$

$\mathbf{S}$ is the $N \times N$ design matrix whose *i*-th row equals $\mathbf{x}_i(\mathbf{X}'\mathbf{G}_i(b,\alpha)\mathbf{X})^{-1}\mathbf{X}'\mathbf{G}_i(b,\alpha)$, where $\mathbf{x}_i$ is the *i*-th row of $\mathbf{X}$. $\hat{\sigma}^2$ is the variance estimator, given by

$$\hat{\sigma}^2 = \frac{1}{N - 2tr[\mathbf{S}] + tr[\mathbf{S'S}]} \sum_{k=1}^{K} (y_i - \mathbf{x}_i \hat{\boldsymbol{\beta}}_i)^2. \tag{A12}$$

Furthermore, the traces $tr[\mathbf{S}]$ and $tr[\mathbf{S'S}]$ can be calculated without explicitly processing the large $\mathbf{S}$ matrix, by using the following expressions (see Appendix 2):

$$tr[\mathbf{S}] = \sum_i \mathbf{x}_i \left[ \alpha \mathbf{M}^{(0)} + \sum_{p=1}^{P} b^p \mathbf{M}_i^{(p)} \right]^{-1} \mathbf{x'}_i g_{i,i}(b, \alpha), \tag{A13}$$

$$tr[\mathbf{S'S}] = \sum_i \mathbf{x}_i \left[ \alpha \mathbf{M}^{(0)} + \sum_{p=1}^{P} b^p \mathbf{M}_i^{(p)} \right]^{-1}$$
$$\left( \alpha^2 \mathbf{M}^{(0)} + 2\alpha \sum_{p=1}^{P} b^p \mathbf{M}_i^{(p)} + \sum_{p=1}^{P} b^{2p} \mathbf{M}_i^{(2p)} \right) \left[ \alpha \mathbf{M}^{(0)} + \sum_{p=1}^{P} b^p \mathbf{M}_i^{(p)} \right]^{-1} \mathbf{x'}_i, \tag{A14}$$

where $\mathbf{M}_i^{(2p)} = \mathbf{M}_i^{(p)} \mathbf{M}_i^{(p)}$ is evaluated before the iterations for the AICc minimization. The most challenging additional CP required for the AICc minimization is the product of the $K \times K$ matrices in $tr[\mathbf{S'S}]$. Given $\mathbf{M}_i^{(p)}$ and $\mathbf{M}^{(0)}$ in step (I), the complexity is independent of $N$. Thus, the AICc-based approach achieves a linear-time estimation as well.